# Robust Initial Alignment for SINS/DVL Based on Reconstructed Observation Vectors

Xiang Xu, Jing Gui, Yifan Sun, Yiqing Yao and Tao Zhang

*Abstract*—**Misalignment angle will result in a considerable error for the integration of Doppler Velocity Log (DVL) and of Strapdown Inertial Navigation System (SINS). In this paper, a robust initial alignment method for SINS/DVL is proposed to solve a practical applicable issue, which is that the outputs of DVL are often corrupted by the outliers. Firstly, the alignment principle for SINS/DVL is summarized. Secondly, based on the principle of this alignment method, the apparent velocity model is investigated, and the parameters expression of the apparent velocity model are derived detailed. Using the apparent velocity model, the unknown parameters of the apparent velocity model are estimated by the developed Robust Kalman Filter (RKF), then the reconstructed observation vector, where the outliers are detected and isolated, is reconstructed by the estimated parameters. Based on the reconstructed observation vectors, the initial attitude is determined. Finally, the simulation and field tests are carried out to verify the performance of the proposed method. The test results are shown that the proposed method can detect and isolate the outliers effectively and get better performance than the previous work.**

## NOMENCLATURE

### A. Reference Frames

| | |
|---|---|
| *i*-frame | The Earth-centered-fixed orthogonal reference frame, which is non-rotating during the alignment stage. |
| *n*-frame | Orthogonal reference frame aligned with East-North-Up (ENU) geodetic axes. |
| *n*0-frame | The orthogonal reference frame, which is formed by fixing the *n*-frame at start up in the inertial space, is non-rotating during the alignment stage. |
| *b*-frame | The orthogonal reference frame aligned with inertial measurement unit (IMU) axes. |
| *b*0-frame | The orthogonal reference frame, which is formed by fixing the *b*-frame at startup in the inertial space to the *i*-frame, is non-rotating during the alignment stage. |
| *e*-frame | Earth-centered Earth-fixed (ECEF) orthogonal reference frame; |
| *e*0-frame | The orthogonal reference frame, which is formed by fixing the *e*-frame at startup in the inertial space to the *i*-frame, is non-rotating during the alignment stage. |

### B. Transforms

Transform subscripts denote the reference frame from which the transform operates, and superscripts denote the reference frame to which the transform operates.

| | |
|---|---|
| $\boldsymbol{q}_b^n$ | The quaternion denoting the rotational transformation between the *b*-frame and the *n*-frame. |
| $\boldsymbol{C}_b^n$ | Direction cosine matrix (DCM) denoting the rotational transformation between the *b*-frame and the *n*-frame. |

This work was supported in part by the National Natural Science Foundation of China under Grants 61803278 and 61974156, in part by the National Equipment Pre-Research Foundation of China (61405170207, 61405170308), in part by the Foundation of Key Laboratory of Micro-Inertial Instrument and Advanced Navigation Technology, Ministry of Education, China (SEU-MIAN-201802), the Inertial Technology Key Lab Fund (614250607011709), the Fundamental Research Funds for the Central Universities (2242018K40065), the Foundation of Shanghai Key Laboratory of Navigation and Location Based Services, Key Laboratory Fund for Underwater Information and Control (614221805051809).

Xiang Xu, Jing Gui and Yifan Sun are with the School of Electronic and Information Engineering, Soochow University, Suzhou 215006, China (e-mail: hsianghsu@163.com; jgui716@163.com; syf69123@sina.com;).

Yiqing Yao and Tao Zhang are with the with the Key Laboratory of Micro-Inertial Instrument and Advanced Navigation Technology, Ministry of Education, Southeast University, Nanjing 210096, China (e-mail: yucia@sina.com; 101011356@seu.edu.cn).

## I. INTRODUCTION

STRAPDOWN Inertial Navigation System (SINS) is of a vital importance for vessels, missiles, infantry fighting vehicles, and other military or consuming equipment [1-3]. Before the navigation procedure, it is necessary to carry out the initial alignment procedure for SINS, because the purpose of navigation task is to determine the position of vehicle in the *n*-frame, but the acceleration and angular rate is sensed by SINS in *b*-frame [4, 5]. Thus, the initial alignment procedure is finding the attitude between *n*-frame and *b*-frame.

To implement the initial alignment procedure efficiently, many researchers were devoted to exploring the novel methods [6-13]. In [6], an attitude determination method was proposed to the analytical calculation for initial attitude. Then, to denoise the measurement of SINS, a wavelet method was investigated in [7]. However, these two methods are just applicable to the static base, if the base is swaying or in-motion, the initial alignment procedure will fail. To address this defect, an inertial frame alignment method was devised in [8]. Based on the inertial frame alignment method, a digital filter method was proposed for filtering the observation vectors, this process can reduce the noise of the observation vectors and improve the alignment performance [9-11]. However, the attitude determination in [9-11] is not optimal. To improve the performance of attitude determination method, an optimization-base method was investigated [12, 13], and the initial alignment procedure transforms into a continuous attitude determination problem using infinite observation vectors. However, these methods cannot finish the initial alignment when the vehicle is in-motion. And, it is noted that in-motion initial alignment procedure is of vital importance for some special application scenarios, such as emergent start-up or start up SINS for a vehicle in highway.

In this respect, how to finish the initial alignment when vehicle is in-motion become a hot topic of SINS technology [14-24]. In [14, 15], a slide window method for selecting the length of observation method was developed, and the outputs of GPS are used as external information. Then, a velocity and position integration formula for SINS was devised to finish the in-motion initial alignment with the *n*-frame velocity of DGPS [16]. To extend the application of this method, an odometer-aided in-motion alignment method was investigated in [17-19]. Meanwhile, for application of autonomous underwater vehicle, a DVL-aided in-motion initial alignment method was proposed in [20-24]. Different from the GPS-aided in-motion initial alignment, the observation vectors of DVL-aided methods were re-derived, and the real-time position and backtracking position of the vehicle, which are not obtained from the external observable information, were calculated by [20] and [23], respectively. Although, many useful methods were proposed in recent years, but a practical issue is ignored in almost all existing previous works, that is the aided information, such as position and velocity of GPS, velocity of odometer and DVL, always contain the outliers. These non-Gaussian noises will corrupt the performance of initial alignment.

In this paper, we focus on solving this practical issue for SINS/SVL integrated navigation system. Firstly, the initial alignment model for SINS/DVL is derived, and the sampling rate disparity between SINS and DVL is considered for discrete calculation. Inspired by the apparent velocity model of the observation and reference vectors, a parameter model of the observation and reference vectors are derived in detail. Using the Huber's M-estimation method, an RKF for parameter identification is investigated. The contributions of this paper are to improve the robustness of initial alignment, and make the proposed method more suitable for the practical applications.

The rest of this paper is organized as follows. Section II mathematically formulates the general form of the initial alignment for SINS/DVL. In Section III, a reconstructed method for the calculated observation vectors is investigated. The efficiency of the proposed method is demonstrated through simulation and field tests results in Section IV. Finally, the conclusions are drawn in Section V.

## II. System Description and Problem Statement

In this section, the traditional initial alignment method for SINS/DVL integrated navigation system is described briefly, and then the existing practical issues are introduced.

### *A. System Description*

According to the coordinate transformation method, it has

$$\dot{\boldsymbol{v}}^n = (\boldsymbol{C}_b^n \boldsymbol{v}^b)' = \dot{\boldsymbol{C}}_b^n \boldsymbol{v}^b + \boldsymbol{C}_b^n \dot{\boldsymbol{v}}^b \tag{1}$$

where, $\boldsymbol{v}^n$ is the velocity of the vehicle in *n*-frame. It is noted that if not explicitly stated, quantities in this paper are time dependent. The dependence on $t$ is omitted for brief presentation. The time derivative of $\boldsymbol{C}_b^n$ is given by

$$\dot{\boldsymbol{C}}_b^n = \boldsymbol{C}_b^n [\boldsymbol{\omega}_{nb}^b \times] \tag{2}$$

The velocity kinematic equation expressed in *n*-frame is given by

$$\dot{\boldsymbol{v}}^n = \boldsymbol{C}_b^n \boldsymbol{f}^b - (2\boldsymbol{\omega}_{ie}^n + \boldsymbol{\omega}_{en}^n) \times \boldsymbol{v}^n + \boldsymbol{g}^n \tag{3}$$

where, $\boldsymbol{f}^b$ is the specific force in *b*-frame, and $\boldsymbol{g}^n$ is the gravity vector in *n*-frame. Based on the chain rules of the DCM (Direction Cosine Matrix) and (1), (3) can be derived as

$$\boldsymbol{C}_b^n [\boldsymbol{\omega}_{nb}^b \times] \boldsymbol{v}^b + \boldsymbol{C}_b^n \dot{\boldsymbol{v}}^b = \boldsymbol{C}_{n0}^n \boldsymbol{C}_{b0}^{n0} \boldsymbol{C}_b^{b0} \boldsymbol{f}^b - (2\boldsymbol{\omega}_{ie}^n + \boldsymbol{\omega}_{en}^n) \times \boldsymbol{v}^n + \boldsymbol{g}^n \tag{4}$$

After some manipulation, (4) can be expressed as

$$\boldsymbol{C}_b^{b0} (\boldsymbol{\omega}_{ie}^b + \boldsymbol{\omega}_{ib}^b) \times \boldsymbol{v}^b + \boldsymbol{C}_b^{b0} \dot{\boldsymbol{v}}^b - \boldsymbol{C}_b^{b0} \boldsymbol{f}^b = \boldsymbol{C}_{n0}^{b0} \boldsymbol{C}_n^{n0} \boldsymbol{g}^n \tag{5}$$

Integrating (5), it has

$$\int_0^t [\boldsymbol{C}_b^{b0} (\boldsymbol{\omega}_{ie}^b + \boldsymbol{\omega}_{ib}^b) \times \boldsymbol{v}^b + \boldsymbol{C}_b^{b0} \dot{\boldsymbol{v}}^b - \boldsymbol{C}_b^{b0} \boldsymbol{f}^b] d\tau = \boldsymbol{C}_{n0}^{b0} \int_0^t \boldsymbol{C}_n^{n0} \boldsymbol{g}^n d\tau \tag{6}$$

Define

$$\begin{cases} \boldsymbol{\beta} = \int_0^t [\boldsymbol{C}_b^{b0} (\boldsymbol{\omega}_{ie}^b + \boldsymbol{\omega}_{ib}^b) \times \boldsymbol{v}^b + \boldsymbol{C}_b^{b0} \dot{\boldsymbol{v}}^b - \boldsymbol{C}_b^{b0} \boldsymbol{f}^b] d\tau \\ \boldsymbol{\alpha} = \int_0^t \boldsymbol{C}_n^{n0} \boldsymbol{g}^n d\tau \end{cases} \tag{7}$$

Then the attitude determination model can be developed as

$$\boldsymbol{\beta} = \boldsymbol{C}_{n0}^{b0} \boldsymbol{\alpha} \tag{8}$$

when the vectors in (7) are obtained, the initial DCM $\boldsymbol{C}_{n0}^{b0}$ can be readily calculated by the attitude determination algorithms [12, 13]. It is noted that the sampling rates of SINS and DVL are different with each other, here we denote $\Delta t_s$ as the sampling interval of the SINS, and $\Delta t_d$ as the sampling interval of the DVL, where $\Delta t_d = D\Delta t_s (D \in \mathbb{N})$. The current time instant during the initial alignment is assumed as $t = M\Delta t_d$. Based on these assumptions, the discrete recursive expression of (7) can be derived as follows.

Firstly, by the partial integration, it has

$$\int_0^t \boldsymbol{C}_b^{b0}\dot{\boldsymbol{v}}^b d\tau = \boldsymbol{C}_b^{b0}\boldsymbol{v}^b\Big|_0^t - \int_0^t \boldsymbol{C}_b^{b0}\left[\boldsymbol{\omega}_{ib}^b \times\right]\boldsymbol{v}^b d\tau \tag{9}$$

Then, substituting (9) into (7), $\boldsymbol{\beta}$ can be calculated as

$$\boldsymbol{\beta} = \boldsymbol{C}_b^{b0}\boldsymbol{v}^b - \boldsymbol{v}^b(0) + \int_0^t \boldsymbol{C}_b^{b0}\left[\boldsymbol{\omega}_{ie}^b \times\right]\boldsymbol{v}^b d\tau - \int_0^t \boldsymbol{C}_b^{b0}\boldsymbol{f}^b d\tau \tag{10}$$

where, $\boldsymbol{C}_b^{b0}(0) = \boldsymbol{I}_3$.

The first integral in (10) can be calculated as

$$\begin{aligned}&\int_0^t \boldsymbol{C}_b^{b0}\left[\boldsymbol{\omega}_{ie}^b \times\right]\boldsymbol{v}^b d\tau\\ &= \sum_{k=0}^{M-1}\boldsymbol{C}_{b(t_k)}^{b0}\int_{t_k}^{t_{k+1}}\boldsymbol{C}_{b(\tau)}^{b(t_k)}\boldsymbol{C}_{n(\tau)}^{b(\tau)}\left[\boldsymbol{\omega}_{ie}^n \times\right]\boldsymbol{v}^n d\tau\\ &= \sum_{k=0}^{M-1}\boldsymbol{C}_{b(t_k)}^{b0}\boldsymbol{C}_{n(t_k)}^{b(t_k)}\int_{t_k}^{t_{k+1}}\boldsymbol{C}_{n(\tau)}^{n(t_k)}\left[\boldsymbol{\omega}_{ie}^n \times\right]\boldsymbol{v}^n d\tau\end{aligned} \tag{11}$$

According to the velocity integration formula in [16], the incremental integral in (11) can be approximated as

$$\begin{aligned}&\Delta\boldsymbol{v}^{n(t_k)} = \int_{t_k}^{t_{k+1}}\boldsymbol{C}_{n(\tau)}^{n(t_k)}\left[\boldsymbol{\omega}_{ie}^n \times\right]\boldsymbol{v}^n d\tau\\ &= \left(\frac{\Delta t_d}{2}\boldsymbol{I}_3 + \frac{\Delta t_d^2}{6}\left[\boldsymbol{\omega}_{in}^n \times\right]\right)\left[\boldsymbol{\omega}_{ie}^n \times\right]\boldsymbol{v}^n(t_k) + \left(\frac{\Delta t_d}{2}\boldsymbol{I}_3 + \right.\\ &\left.\frac{\Delta t_d^2}{3}\left[\boldsymbol{\omega}_{in}^n \times\right]\right)\left[\boldsymbol{\omega}_{ie}^n \times\right]\boldsymbol{v}^n(t_{k+1})\end{aligned} \tag{12}$$

where,

$$\begin{cases}\boldsymbol{v}^n(t_k) = \boldsymbol{C}_{b(t_k)}^{n(t_k)}\boldsymbol{v}^b(t_k)\\ \boldsymbol{v}^n(t_{k+1}) = \boldsymbol{C}_{n(t_k)}^{n(t_{k+1})}\boldsymbol{C}_{b(t_k)}^{n(t_k)}\boldsymbol{C}_{b(t_{k+1})}^{b(t_k)}\boldsymbol{v}^b(t_{k+1})\end{cases} \tag{13}$$

$$\begin{cases}\boldsymbol{C}_{b(t_{k+1})}^{b(t_k)} = \boldsymbol{I}_3 + \frac{\sin\|\boldsymbol{\theta}_{ib}^b\|}{\|\boldsymbol{\theta}_{ib}^b\|}\left[\boldsymbol{\Theta}_{ib}^b \times\right] + \frac{1-\cos\|\boldsymbol{\theta}_{ib}^b\|}{\|\boldsymbol{\theta}_{ib}^b\|^2}\left[\boldsymbol{\Theta}_{ib}^b \times\right]^2\\ \boldsymbol{C}_{n(t_{k+1})}^{n(t_k)} = \boldsymbol{I}_3 + \frac{\sin\|\boldsymbol{\theta}_{in}^n\|}{\|\boldsymbol{\theta}_{in}^n\|}\left[\boldsymbol{\Theta}_{in}^n \times\right] + \frac{1-\cos\|\boldsymbol{\theta}_{in}^n\|}{\|\boldsymbol{\theta}_{in}^n\|^2}\left[\boldsymbol{\Theta}_{in}^n \times\right]^2\end{cases} \tag{14}$$

$$\begin{cases}\boldsymbol{\Theta}_{ib}^b = \sum_{l=1}^{D}\boldsymbol{\omega}_{ib,l}^b\Delta t_s\\ \boldsymbol{\Theta}_{in}^n \approx \boldsymbol{\omega}_{in}^n\Delta t_d\end{cases} \tag{15}$$

where, $\Delta t_d = t_{k+1} - t_k$. It is noted that $\boldsymbol{\omega}_{ib,l}^b$ is the angular rate of the gyroscope measurement during $t_k$ to $t_{k+1}$, and $\boldsymbol{\omega}_{in}^n = \boldsymbol{\omega}_{ie}^n + \boldsymbol{\omega}_{en}^n$ is the angular rate of $n$-frame with respect to $i$-frame. In this paper, when the vehicle is in the procedure of initial alignment, the moving velocity is low. Moreover, the duration of the initial alignment procedure is short. Thus, the magnitude of $\boldsymbol{\omega}_{en}^n$ is much smaller than $\boldsymbol{\omega}_{ie}^n$. Thus, we use $\boldsymbol{\omega}_{ie}^n$ instead of $\boldsymbol{\omega}_{in}^n$ to calculate $\boldsymbol{\Theta}_{in}^n$.

The second integral in (10) can be calculated as

$$\int_0^t \boldsymbol{C}_b^{b0}\boldsymbol{f}^b d\tau = \sum_{k=0}^{M-1}\boldsymbol{C}_{b(t_k)}^{b0}\int_{t_k}^{t_{k+1}}\boldsymbol{C}_{b(\tau)}^{b(t_k)}\boldsymbol{f}^b(\tau)d\tau \tag{16}$$

The incremental integral in (16) can be approximated using the two-sample correction by [16]

$$\begin{aligned}&\Delta\boldsymbol{v}^{b(t_k)} = \int_{t_k}^{t_{k+1}}\boldsymbol{C}_{b(\tau)}^{b(t_k)}\boldsymbol{f}^b(\tau)d\tau\\ &= \Delta\boldsymbol{v}_1 + \Delta\boldsymbol{v}_2 + \frac{1}{2}(\Delta\boldsymbol{\vartheta}_1 + \Delta\boldsymbol{\vartheta}_2) \times (\Delta\boldsymbol{v}_1 + \Delta\boldsymbol{v}_2) +\\ &\frac{2}{3}(\Delta\boldsymbol{\vartheta}_1 \times \Delta\boldsymbol{v}_2 + \Delta\boldsymbol{v}_1 \times \Delta\boldsymbol{\vartheta}_2)\end{aligned} \tag{17}$$

Substituting (11), (12), (16) and (17) into (10), and using the recursive form to describe the results

$$\boldsymbol{\beta}_M = \boldsymbol{C}_{b(M)}^{b0}\boldsymbol{v}^b(M) - \boldsymbol{v}^b(0) + \boldsymbol{\beta}'_M \tag{18}$$

where,

$$\begin{aligned}&\boldsymbol{\beta}'_M = \boldsymbol{\beta}'_{M-1} + \boldsymbol{C}_{b(M-1)}^{b0}\boldsymbol{C}_{n(M-1)}^{b(M-1)}\Delta\boldsymbol{v}^{n(M-1)}\\ &-\boldsymbol{C}_{b(M-1)}^{b0}\Delta\boldsymbol{v}^{b(M-1)}\end{aligned} \tag{19}$$

According to the aforementioned derivation, the observation vector $\boldsymbol{\beta}_M$ can be calculated by (18) and (19) iteratively. And, the reference vector $\boldsymbol{\alpha}$, which is shown in (7), can be calculated as the iterative form

$$\boldsymbol{\alpha}_M = \boldsymbol{\alpha}_{M-1} + \boldsymbol{C}_{n(M-1)}^{n0}\left(\Delta t_d\boldsymbol{I}_3 + \frac{\Delta t_d^2}{2}\left[\boldsymbol{\omega}_{in}^n \times\right]\right)\boldsymbol{g}^n \tag{20}$$

It is noted that the DCM $\boldsymbol{C}_{n0}^{b0}$ is calculated at the updated interval of the DVL.

Based on the calculated vectors, the initial attitude can be determined by the optimized-based attitude determination algorithm [13]

$$\boldsymbol{K}_M = \boldsymbol{K}_{M-1} + ([\boldsymbol{\beta}_M\otimes] - [\boldsymbol{\alpha}_M\odot])^{\mathrm{T}}([\boldsymbol{\beta}_M\otimes] - [\boldsymbol{\alpha}_M\odot]) \tag{21}$$

where,

$$\begin{cases}[\boldsymbol{\beta}_M\otimes] = \begin{bmatrix}\boldsymbol{0} & -\boldsymbol{\beta}_M^{\mathrm{T}}\\ \boldsymbol{\beta}_M & [\boldsymbol{\beta}_M \times]\end{bmatrix}\\ [\boldsymbol{\alpha}_M\odot] = \begin{bmatrix}\boldsymbol{0} & -\boldsymbol{\alpha}_M^{\mathrm{T}}\\ \boldsymbol{\alpha}_M & -[\boldsymbol{\alpha}_M \times]\end{bmatrix}\end{cases} \tag{22}$$

The attitude quaternion $\boldsymbol{q}_{n0}^{b0}$ at time instant $M$ can be extracted from the matrix $\boldsymbol{K}_M$. It is the normalized eigenvector corresponding to the minimum eigenvalue of $\boldsymbol{K}_M$. Then, the DCM $\boldsymbol{C}_{n0}^{b0}$ can be calculated by the quaternion $\boldsymbol{q}_{n0}^{b0}$. And the real-time attitude of the vehicle $\boldsymbol{C}_b^n$ can be obtained by the chain rule of DCM according to the $\boldsymbol{C}_{n0}^{b0}$ and (14).

*B. Problem Statement*

The Doppler velocity is obtained through measuring the Doppler shifts of the signals radiated by three or more non-coplanar radio beams off a surface [21]. The measurement accuracy varies with frequency, instrument design and external environment, such as water temperature and depth and so on. There will be inevitable outliers, which are contained in the outputs of DVL. Supposing the misalignment angles between DVL and SINS and the scale factor errors of DVL are corrected before the initial alignment procedure. Then the model of the

DVL measurements can be given by

$$\tilde{\boldsymbol{v}}^b(M) = \boldsymbol{v}^b(M) + \delta\boldsymbol{v}_o^b \tag{23}$$

where, $\delta\boldsymbol{v}_o^b = \delta\boldsymbol{v}^b + \mu_i\mathbf{e}_i$ , $\delta\boldsymbol{v}^b \sim N(\mathbf{0},\ \sigma^2\boldsymbol{I}_3)$ is the measurement noises of DVL, which is zero mean and variance $\sigma^2\boldsymbol{I}_3$. For the alternative hypothesis, the error vector is $\mathbf{e}_i(i = 1,2,3)$, such that only the $i$th element is 1. And the magnitude of the error is $\mu_i$, when the quantity $\mu_i$ is nonzero, the measurement $\tilde{\boldsymbol{v}}^b(M)$ is called an outlier.

Substituting (23) into (18), and ignoring the noises of inertial sensors, the calculated observation vector with the outputs of SINS and DVL can be expressed as

$$\tilde{\boldsymbol{\beta}}_M = \boldsymbol{C}_{b(M)}^{b0}\boldsymbol{v}^b(M) - \boldsymbol{v}^b(0) + \boldsymbol{\beta}'_M + \boldsymbol{C}_{b(M)}^{b0}\delta\boldsymbol{v}_o^b \tag{24}$$

When $\mu_i$ is nonzero, the calculated observation vector $\tilde{\boldsymbol{\beta}}_M$ will be contaminated by the outliers. Since the initial attitude is determined by the calculated observation vector, the outliers will decrease the accuracy of the attitude determination algorithm.

To detect the outliers, many researchers are devoted to developing the efficient methods [25-28]. Such outlier's detection is mainly based on the integration filtering through robust algorithm. Unfortunately, the outliers cannot be detected by the robust estimation algorithm with the analytical attitude determination method, which is shown in (21).

Inspired by the apparent motion of the reference vector of $\boldsymbol{\alpha}_M$, the parameter model of the vector $\boldsymbol{\beta}_M$ is investigated. In this respect, we proposed a parameter identification method to reconstruct the observation vector $\boldsymbol{\beta}_M$ . During the reconstruction procedure, an RKF algorithm based on the Huber's M-estimation principle is used to detect the outliers and isolate them. Thus, the proposed method is more robust than the traditional method.

## III. Reconstructed Methods for Observation Vector

This section derives the parameter model of the observation vector, which is called the apparent velocity model in this paper. According to the derived apparent velocity model, the unknown constant parameters of the model are estimated by the RKF. Moreover specifically, the outliers contained in the calculated observation vector $\tilde{\boldsymbol{\beta}}_M$ are detected and isolated. Finally, the new vector, which is reconstructed by the estimated parameters, is used to determining the initial attitude.

### *A. Apparent Velocity Model*

Based on the chain rule of the DCM, (7) can be rewritten as

$$\boldsymbol{\alpha} = \int_0^t \boldsymbol{C}_{e0}^{n0}\boldsymbol{C}_e^{e0}\boldsymbol{C}_n^e\boldsymbol{g}^n d\tau \tag{25}$$

where, $\boldsymbol{C}_{e0}^{n0}$ is related to the latitude and longitude at the very start of the initial alignment. $\boldsymbol{C}_e^{e0}$ is related to the angular rate and the alignment time $t$. $\boldsymbol{C}_n^e$ is related to the latitude and longitude at current moment. Since the short alignment time and low moving velocity of the vehicle, the DCM $\boldsymbol{C}_n^e$ can be approximated to the DCM $\boldsymbol{C}_{n0}^{e0}$.

Thus, (25) can be calculated as

$$\boldsymbol{\alpha} = \boldsymbol{\Phi}\boldsymbol{\Gamma}(t) \tag{26}$$

where,

$$\begin{cases} \boldsymbol{\Phi} = \begin{bmatrix} \frac{g\,\mathrm{co}}{\omega_{ie}} & 0 & 0 & -\frac{g\cos L}{\omega_{ie}} \\ 0 & \frac{g\cos L\sin L}{\omega_{ie}} & -g\cos L\sin L & 0 \\ 0 & -\frac{g\cos^2 L}{\omega_{ie}} & -g\sin^2 L & 0 \end{bmatrix} \\ \boldsymbol{\Gamma}(t) = [\cos(\omega_{ie}t) \quad \sin(\omega_{ie}t) \quad t \quad 1]^{\mathrm{T}} \end{cases} \tag{27}$$

where, $g$ is the magnitude of the gravity at the latitude $L$. $\omega_{ie}$ is the angular rate of the Earth, it is a constant value during the whole alignment procedure. With consideration of the short alignment time, the latitude $L$ can be approximated as the constant value during the alignment procedure. Therefore, the matrix $\boldsymbol{\Xi}$ is a constant matrix.

Based on the parameter model of the reference vector, the parameter model of the observation vector $\boldsymbol{\beta}$ can be calculated by (8), because the DCM $\boldsymbol{C}_{n0}^{b0}$ is also a constant matrix during the whole alignment procedure

$$\boldsymbol{\beta} = \boldsymbol{\Xi}\boldsymbol{\Gamma}(t) \tag{28}$$

where, $\boldsymbol{\Xi} = \boldsymbol{C}_{n0}^{b0}\boldsymbol{\Phi}$.

Substituting (28) into (24)

$$\tilde{\boldsymbol{\beta}}_M = \boldsymbol{\Xi}\boldsymbol{\Gamma}(M) + \boldsymbol{C}_{b(M)}^{b0}\delta\boldsymbol{v}_o^b \tag{29}$$

The parameter model (29) is the apparent velocity model of the calculated observation vector. It has been pointed out that, the change of the vector $\tilde{\boldsymbol{\beta}}_M$ is regular. And if the parameter matrix $\boldsymbol{\Xi}$, which is denoted as $\hat{\boldsymbol{\Xi}}$, can be estimated, the optimal vector $\hat{\boldsymbol{\beta}}_M$ can be reconstructed by $\hat{\boldsymbol{\Xi}}\boldsymbol{\Gamma}(M)$. Moreover, the robust algorithm can be used to the parameter estimation method, thus the outliers can be detected and isolated [29,30]. The detailed derivation is depicted in the ensuing subsection.

### *B. Parameter Identification by RKF*

Based on the aforementioned analysis, the key to implement the robust initial alignment for SINS/DVL is to obtain the parameter matrix $\hat{\boldsymbol{\Xi}}$. Moreover, three channels of $\tilde{\boldsymbol{\beta}}_M$ are independent with each other. To simplify the analysis, the third channel is taken as an example to construct the filtering model, which is given by

$$\begin{cases} \boldsymbol{\Xi}_{z,M} = \boldsymbol{\Xi}_{z,M-1} + \boldsymbol{w}_{z,M-1} \\ \tilde{\beta}_{z,M} = \boldsymbol{\Xi}_{z,M}\boldsymbol{\Gamma}(M) + \delta v_{o,z}^{b0}(M) \end{cases} \tag{30}$$

where, $\boldsymbol{\Xi}_{z,M}$ is the third row of the matrix $\boldsymbol{\Xi}_M$, and $\boldsymbol{\Xi}_{z,M}\boldsymbol{\Gamma}(M) = \left(\boldsymbol{\Gamma}(M)\right)^{\mathrm{T}}\left(\boldsymbol{\Xi}_{z,M}\right)^{\mathrm{T}}$.

In this paper, the parameter identification process is solved by the Huber's M-estimation method [25, 26], which has been widely used to the practical system [30]. Next, the detailed filtering process has been derived.

(1) Time update

$$\hat{\boldsymbol{\Xi}}_{z,M|M-1} = \hat{\boldsymbol{\Xi}}_{z,M-1|M-1} \tag{31}$$

$$\boldsymbol{P}_{z,M|M-1} = \boldsymbol{P}_{z,M-1|M-1} + \boldsymbol{Q}_{z,M-1|M-1} \tag{32}$$

where, $\boldsymbol{Q}_{z,M-1|M-1} = \mathrm{E}\left[\left(\boldsymbol{w}_{z,M-1}\right)^{\mathrm{T}}\boldsymbol{w}_{z,M-1}\right]$.

(2) Robust filtering

Define

$$\zeta_{z,M} = \left(\sqrt{R_{z,M|M}}\right)^{-1}\left(\tilde{\beta}_{z,M} - \hat{\boldsymbol{\Xi}}_{z,M|M-1}\boldsymbol{\Gamma}(M)\right) \tag{33}$$

and, $R_{z,M|M} = \mathrm{E}\left[\left(\delta v_{o,z}^{b0}(M)\right)^{\mathrm{T}}\delta v_{o,z}^{b0}(M)\right]$.

Choosing the Huber's score function as

$$\rho\left(\zeta_{z,M}\right) = \begin{cases}\gamma\left|\zeta_{z,M}\right| - \frac{\gamma^2}{2} & \left|\zeta_{z,M}\right| \ge \gamma \\ \frac{{\zeta_{z,M}}^2}{2} & \left|\zeta_{z,M}\right| < \gamma\end{cases} \tag{34}$$

where, $\gamma$ is chosen to give the desired efficiency at the Gaussian model.

Let $\left.\frac{\partial\rho(\zeta_{z,M})}{\partial\Xi_{z,M}}\right|_{\Xi_{z,M}=\hat{\Xi}_{z,M|M}} = 0$, the weight function can be solved as

$$\varpi\left(\zeta_{z,M}\right) = \begin{cases}\frac{\gamma\,\mathrm{sgn}(\zeta_{z,M})}{\zeta_{z,M}} & \left|\zeta_{z,M}\right| \ge \gamma \\ 1 & \left|\zeta_{z,M}\right| < \gamma\end{cases} \tag{35}$$

where, $\mathrm{sgn}(\cdot)$ represent a symbol function.

Based on the aforementioned derivation, the reconstructed measurement can be calculated as

$$\breve{\beta}_{z,M} = \hat{\boldsymbol{\Xi}}_{z,M|M-1}\boldsymbol{\Gamma}(M) + \varpi\left(\tilde{\zeta}_{z,M}\right)\left(\tilde{\beta}_{z,M} - \hat{\boldsymbol{\Xi}}_{z,M|M-1}\boldsymbol{\Gamma}(M)\right) \tag{36}$$

(3) Measurement update

$$\boldsymbol{G}_{z,M} = \boldsymbol{P}_{z,M|M-1}\boldsymbol{\Gamma}(M)\left(\left(\boldsymbol{\Gamma}(M)\right)^{\mathrm{T}}\boldsymbol{P}_{z,M|M-1}\boldsymbol{\Gamma}(M) + R_{z,M|M}\right)^{-1} \tag{37}$$

$$\hat{\boldsymbol{\Xi}}_{z,M|M} = \hat{\boldsymbol{\Xi}}_{z,M|M-1} + \left(\boldsymbol{G}_{z,M}\right)^{\mathrm{T}}\left(\breve{\beta}_{z,M} - \hat{\boldsymbol{\Xi}}_{z,M|M-1}\boldsymbol{\Gamma}(M)\right) \tag{38}$$

$$\boldsymbol{P}_{z,M|M} = \boldsymbol{P}_{z,M|M-1} - \boldsymbol{G}_{z,M}\left(\boldsymbol{\Gamma}(M)\right)^{\mathrm{T}}\boldsymbol{P}_{z,M|M-1} \tag{39}$$

The other two channels, which are $x$-axis and $y$-axis, can also be filtered by the aforementioned method.

### *C. Vector Reconstruction and Algorithm Summarization*

When the parameter matrix $\hat{\boldsymbol{\Xi}}_{M|M}$ has been estimated, the non-outliers vector can be calculated as

$$\hat{\boldsymbol{\beta}}_M = \hat{\boldsymbol{\Xi}}_{M|M}\boldsymbol{\Gamma}(M) \tag{40}$$

Substituting (40) into (21), the non-outliers matrix $\hat{\boldsymbol{K}}_M$ can be obtained. By extracting the normalized eigenvector corresponding to the smallest eigenvalue of $\hat{\boldsymbol{K}}_M$, the initial attitude quaternion $\boldsymbol{q}_{n0}^{b0}$ can be obtained. Based on the chain rules of the DCM, the real-time attitude of the vehicle can be calculated.

At this point, the robust initial alignment for SINS/DVL has been implemented. For clarity, the algorithm procedure has summarized in Table I.

TABLE I
ROBUST INITIAL ALIGNMENT FOR SINS/DVL

| |
|---|
| **Initialization:** $M = 1$, $\hat{\boldsymbol{K}}_0 = \boldsymbol{0}$, $\boldsymbol{C}_b^{b0}(0) = \boldsymbol{C}_n^{n0}(0) = \boldsymbol{I}_3$, $\boldsymbol{\alpha}_0 = \boldsymbol{\beta}_0' = \boldsymbol{0}$ |
| **Inputs:** $\{\boldsymbol{f}^b\}_{k=1}^M$, $\{\boldsymbol{\omega}_{ib}^b\}_{k=1}^M$, $\{\tilde{\boldsymbol{v}}^b\}_{k=1}^M$ |
| **for** $k = 1,2,3,\cdots$ **do** |
| Calculate $\boldsymbol{C}_{n(k)}^{n0}$ and $\boldsymbol{C}_{b(k)}^{b0}$ using (14) and (15) |
| **if** DVL outputs are available |
| Calculate the observation vector according to |
| $\tilde{\boldsymbol{\beta}}_M = \boldsymbol{C}_{b(M)}^{b0}\tilde{\boldsymbol{v}}^b(M) - \tilde{\boldsymbol{v}}^b(0) + \boldsymbol{\beta}_M'$ |
| $\boldsymbol{\beta}_M'$ is calculated by (12), (17), and (19) |
| Calculate the reference vector according to |
| $\boldsymbol{\alpha}_M = \boldsymbol{\alpha}_{M-1} + \boldsymbol{C}_{n(M-1)}^{n0}\left(\Delta t_d\boldsymbol{I}_3 + \frac{\Delta t_d^2}{2}[\boldsymbol{\omega}_{in}^n\times]\right)\boldsymbol{g}^n$ |
| Construct $\boldsymbol{\Gamma}(M) = [\cos(\omega_{ie}M\Delta t_d)\quad \sin(\omega_{ie}M\Delta t_d)\quad \Delta t_d\quad 1]^{\mathrm{T}}$ |
| Calculate $\hat{\boldsymbol{\Xi}}_{j,M\|M-1}(j = x,y,z)$ and $\boldsymbol{P}_{j,M\|M-1}$ using (31) and (32) |
| Calculate $\tilde{\zeta}_{j,M}$, and $\varpi(\tilde{\zeta}_{j,M})$ using (33) and (35) |
| Reconstruct $\breve{\beta}_{z,M}$ using (36) |
| Measurement update for the RKF using (37)-(39) |
| Reconstruct the observation vector $\hat{\boldsymbol{\beta}}_M$ using (40) |
| Using $\hat{\boldsymbol{\beta}}_M$ and $\boldsymbol{\alpha}_M$ to calculate the non-outliers matrix $\hat{\boldsymbol{K}}_M$ |
| Extract $\boldsymbol{q}_{n0}^{b0}$ from $\hat{\boldsymbol{K}}_M$, and transform to $\boldsymbol{C}_{n0}^{b0}$ |
| $M = M + 1$ |
| **end if** |
| Calculate the current attitude according to |
| $\boldsymbol{C}_b^n(k) = \boldsymbol{C}_{n0}^{n(k)}\boldsymbol{C}_{b0}^{n0}\boldsymbol{C}_{b(k)}^{b0}$ |
| **end for** |

## IV. SIMULATION AND FIELD TESTS

In this section, simulation and field tests are carried out to verify the performance of the proposed method. In the simulation test, the vehicle moves along the well-define trajectory. In addition, all the actual information of the movement states is well known. Meanwhile, the field test is designed to verify the performance of the proposed method in the real-time system. Specifically, the following four initial alignment schemes are evaluated for comparison.

*Scheme 1*: the current popular method, which is proposed in [21], with outliers in the outputs of DVL

*Scheme 2*: the method proposed in this paper with the outliers in the outputs of DVL

*Scheme 3*: the current popular method proposed in [21] without outliers in the outputs of DVL

*Scheme 4*: the proposed method in this paper without outliers in the outputs of DVL

### *A. Simulation Test*

This subsection is devoted to numerically examine the proposed method in this paper. We carried out scenarios with S-turn motion on the horizontal plane. Fig. 1 shows the movement attitude and the trajectory of the vehicle.

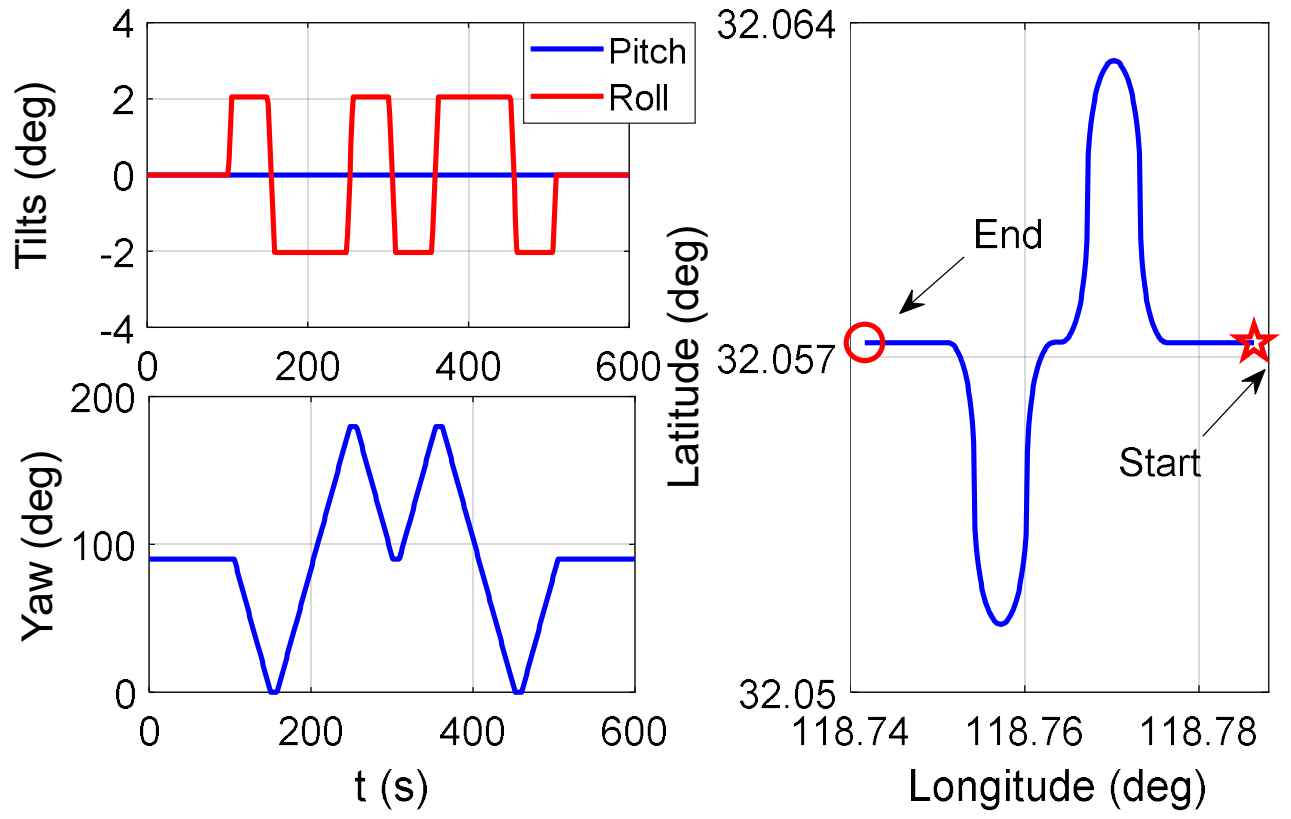

Fig. 1. Curves of the movement attitude and trajectory.

The vehicle under investigation is located at latitude 32.057313 °N and longitude 118.786365 °E. The SINS is equipped with a triad gyroscopes (drift 0.02 °/h, noise $0.005°/\sqrt{\mathrm{h}}$) and accelerometers (bias 50μg, noise $50\mu\mathrm{g}/\sqrt{\mathrm{Hz}}$). The sampling rate of SINS is 200Hz. The outliers corrupted the measurement of DVL are generated according to

$$\delta \boldsymbol{v}_o^{b0}(M) \sim \begin{cases} N(\mathbf{0}, & (0.1)^2\, \boldsymbol{I}_3) & \mathrm{w.p.}\ 0.98 \\ N(\mathbf{0}, & (30)^2\, \boldsymbol{I}_3) & \mathrm{w.p.}\ 0.02 \end{cases} \tag{41}$$

The sampling rate of DVL is 1Hz. And, the measurement velocity of DVL outputs with outliers is shown in Fig. 2.

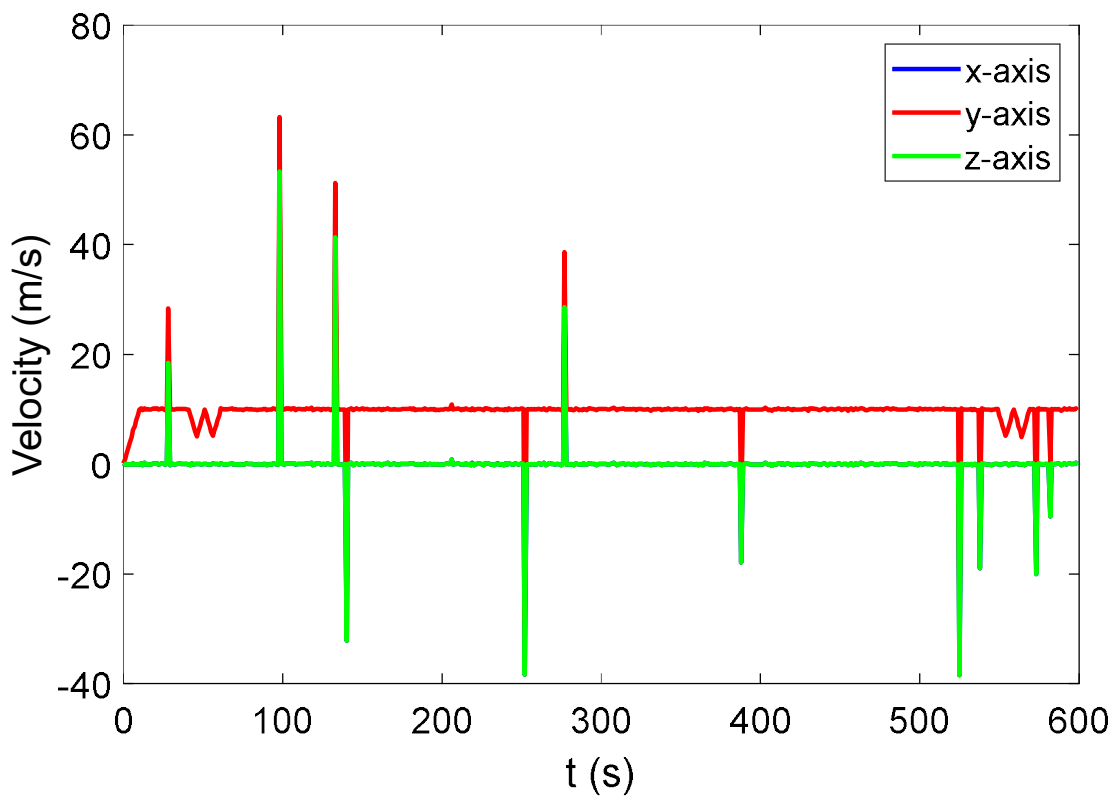

Fig. 2. Curves the measurement velocity of DVL outputs.

The initialization of RKF for parameter identification is set as: $\gamma = 1.345$, $R_{j,M|M} = (0.1\mathrm{m/s})^2$, $\boldsymbol{Q}_{j,M-1|M-1} = (10^{-3}\mathrm{m/s})^2 \boldsymbol{I}_4$, $\boldsymbol{P}_{j,0|0} = (10^5\mathrm{m/s})^2 \boldsymbol{I}_4$, $\hat{\boldsymbol{\Xi}}_{j,0|0} = \mathbf{0}$, where $j = x, y, z$.

The initial alignment results are shown in Fig. 3 to Fig. 6. In Fig. 3, the calculated observation vectors and the reconstructed observation vectors, which are used to attitude determination for each method, are depicted. It is obviously that the calculated observation vectors of Scheme 1 are corrupted by the outliers of DVL outputs. These outliers will decrease the performance of the initial alignment straightforwardly, because the matrix $\boldsymbol{K}_M$ is constructed by the observation vectors. While, in the advantage of the parameter identification method with RKF, the outliers are eliminated from the reconstructed observation vector of Scheme 2. Moreover specifically, the random noises, which are also contained in the DVL outputs, are filtered out by the reconstructed method. It can be found in the corresponding enlarged view, which is shown in the right subplots in Fig. 3. Thus, the reconstructed observation vectors are more accurate than the calculated observation vectors, and the alignment errors are verified this conclusion.

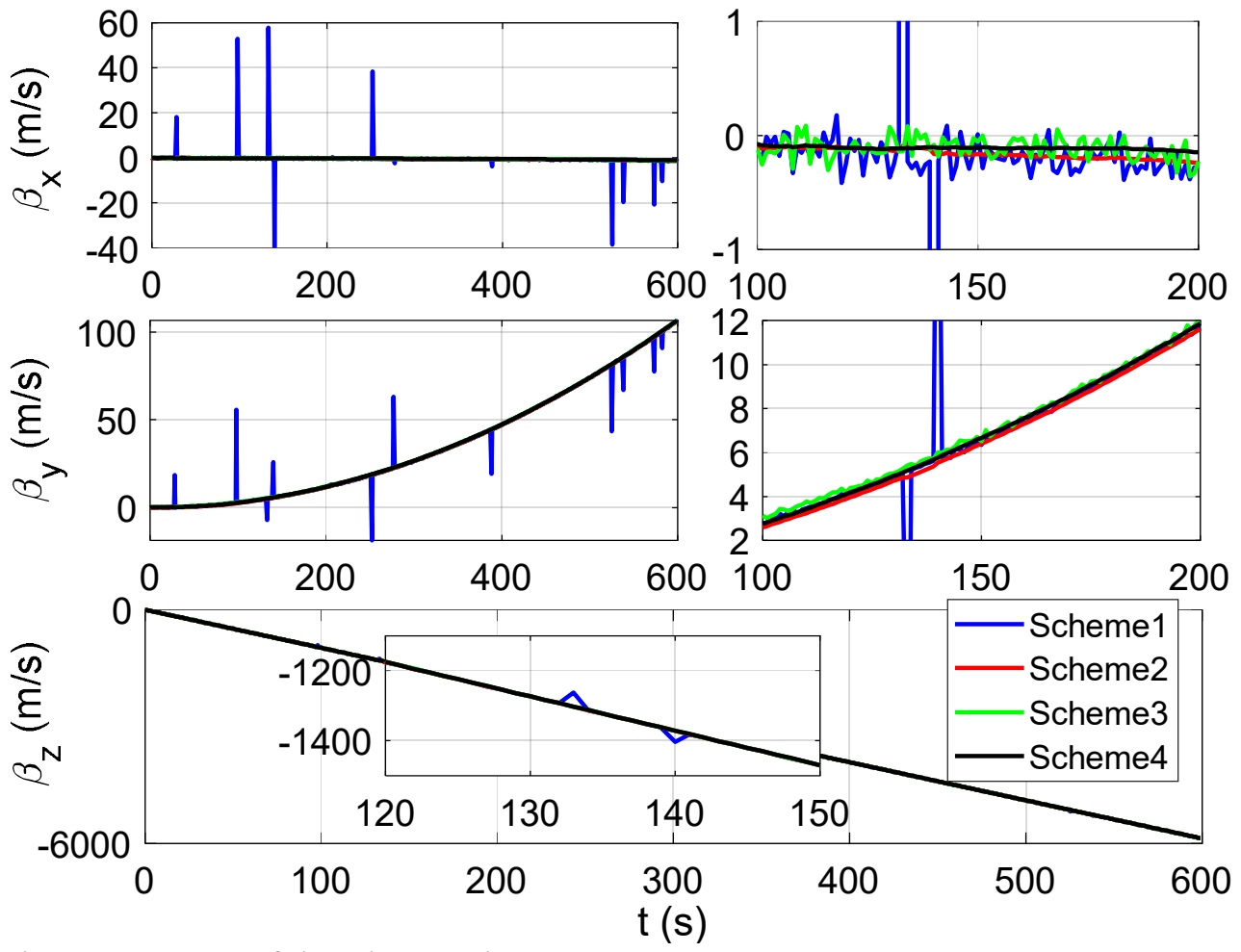

Fig. 3. Curves of the observation vectors.

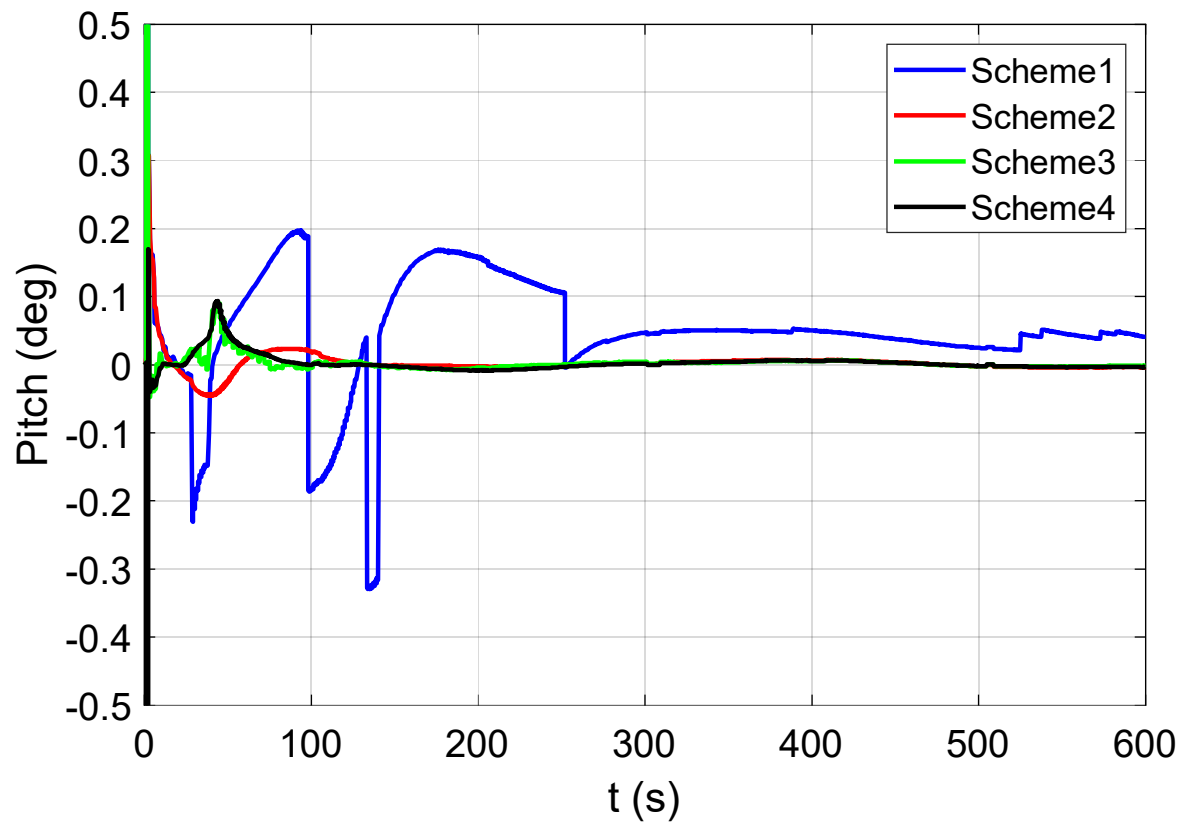

Fig. 4. Alignment errors of pitch by different schemes.

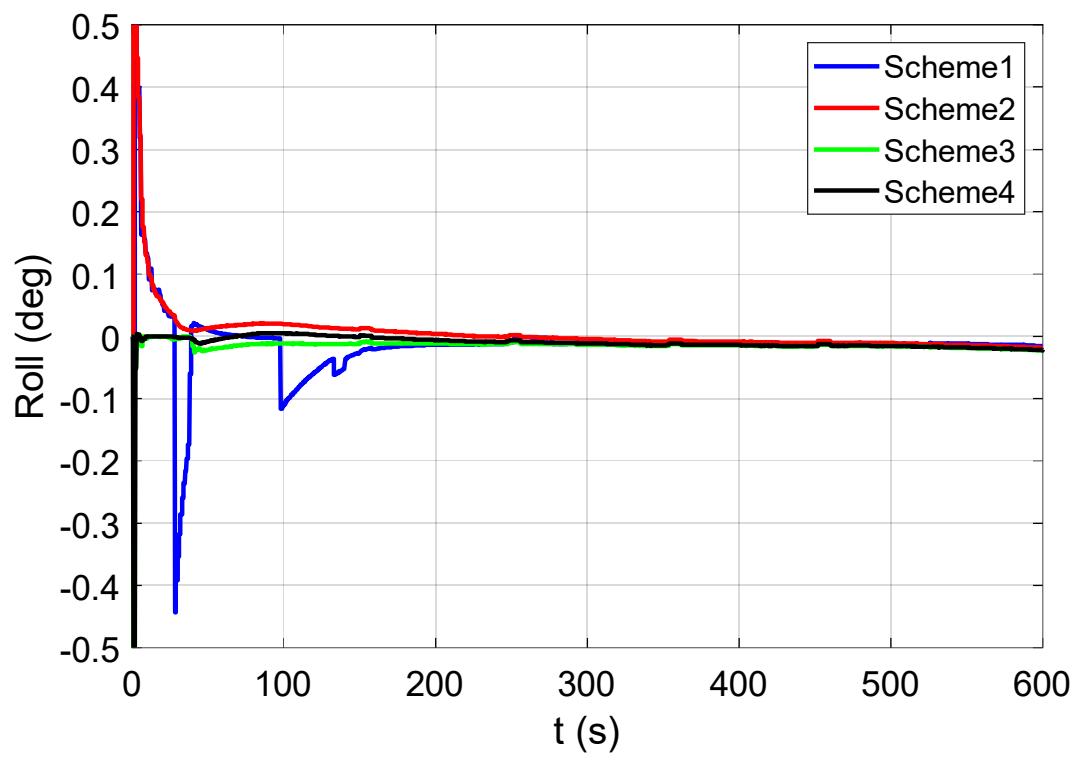

Fig. 5. Alignment errors of roll by different schemes.

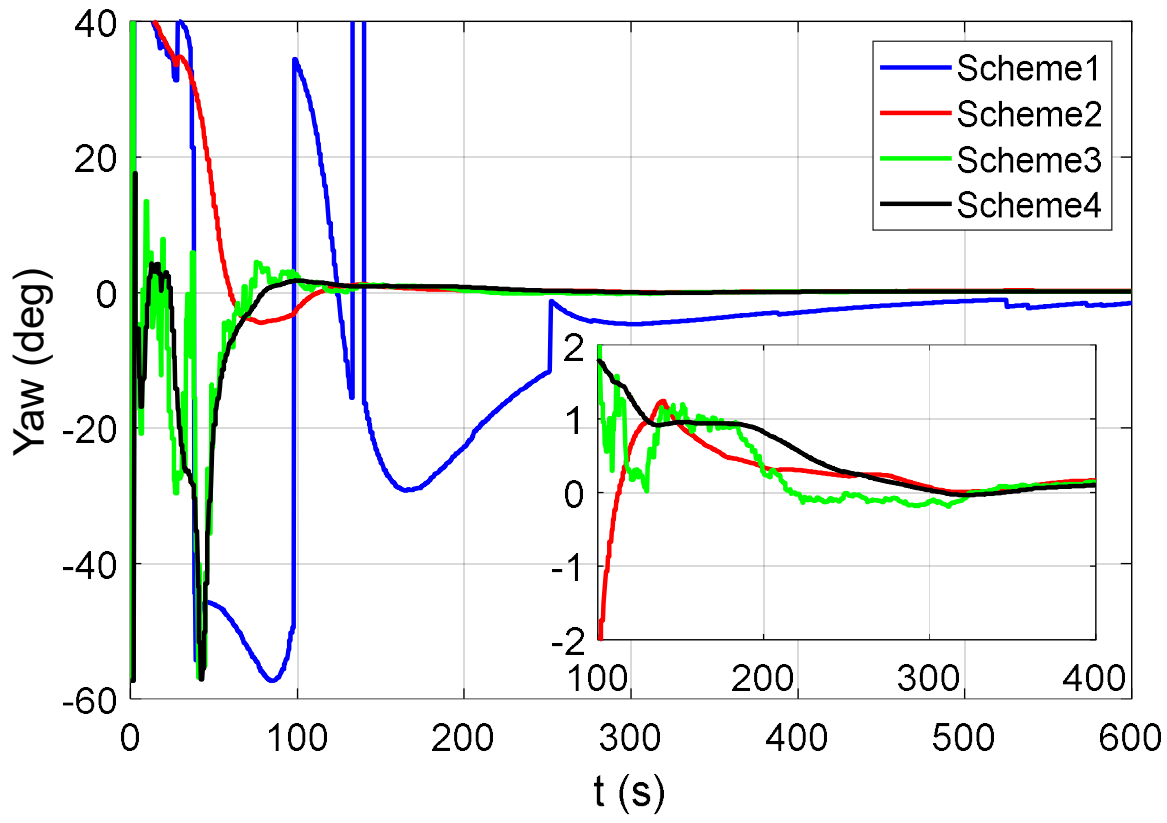

Fig. 6. Alignment errors of yaw by different schemes.

Fig. 4 to 6 are the alignment errors of pitch, roll and yaw respectively. Comparing scheme 1 and scheme 2, it can be found that the outliers are corrupted the alignment results of Scheme 1. However, they have no effect on scheme 2. Comparing scheme 3 and 4, it can be found that the alignment results of scheme 4 are smoother than scheme 3. When the alignment process lasts for 200s, the errors of pitch and roll of scheme 2 to 4 are less than 0.01 degree. The error of yaw of scheme 2 to 4 are less than 1 degree, and it is decrease with alignment time. It is noted that the alignment performance of scheme 2, which aided velocity is corrupted by the outliers, is equivalent with scheme 3 and 4, which aided velocity is not corrupted by the outlier. Moreover, this conclusion is verified the efficiency of the proposed method.

## *B. Field Test*

The field test was carried out on the vehicle, as shown in Fig.7. The PHINS, which is produced by the iXBlue Corporation, is utilized as the reference system [31]. The roll and pitch dynamic accuracy of PHINS is 0.01degree (RMS), and the heading accuracy of PHINS is 0.01sec*L* degree with GPS aiding. Specifically, the SINS and PHINS are installed on the surface of the steel plate. Moreover, the misalignment angle between SINS and PHINS was corrected in advance, it is compensated for comparison. The non-outlier outputs of DVL are generated by the velocity of PHINS using coordinate transformation. The navigational computer is devised by our team, and a real-time operating system is embedded in the computer. The GPS receiver is produced by the NovAtel, the BESTVEL log is used to generate the DVL outputs [32]. The outliers contained in the GPS outputs are generated by the covers or multipath effects. The sampling rate of the GPS is 1Hz. The SINS is equipped with a triad fiber optic gyroscopes and quartz flexible accelerometers. The specifications of the inertial sensors, which is adopted in this test, are listed in Table II. The sampling rate of SINS is 200Hz.

The field test was carried out in Nanjing. The movement trajectory is shown in Fig.8, the distance of the total route is 13.188km. The attitude and the DVL measurements are shown in Fig. 9. It can be found that there are outliers contained in the outputs of DVL.

TABLE II
THE SPECIFICATIONS OF THE INERTIAL SENSORS

| Error item | Gyroscope ($x$-, $y$-, $z$-axes) | Accelerometer ($x$-, $y$-, $z$-axes) |
|---|---|---|
| Bias | $\le$ (0.02, 0.02, 0.02) °/h | $\le$ (6, 6, 6)×10$^3$μg |
| Random walk noise | $\le$ (5, 5, 5)×10$^{-3}$°/$\sqrt{h}$ | $\le$ (50, 50, 50) μg/$\sqrt{Hz}$ |

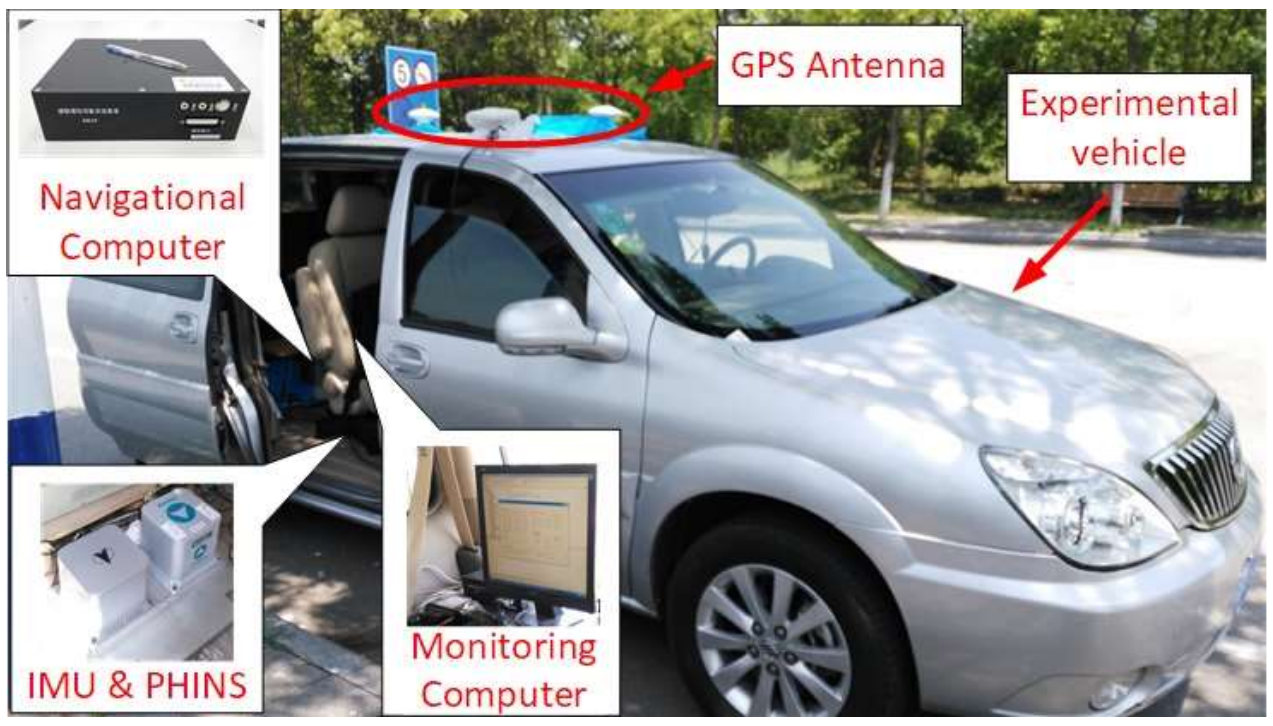

Fig. 7. Field test vehicle and equipment.

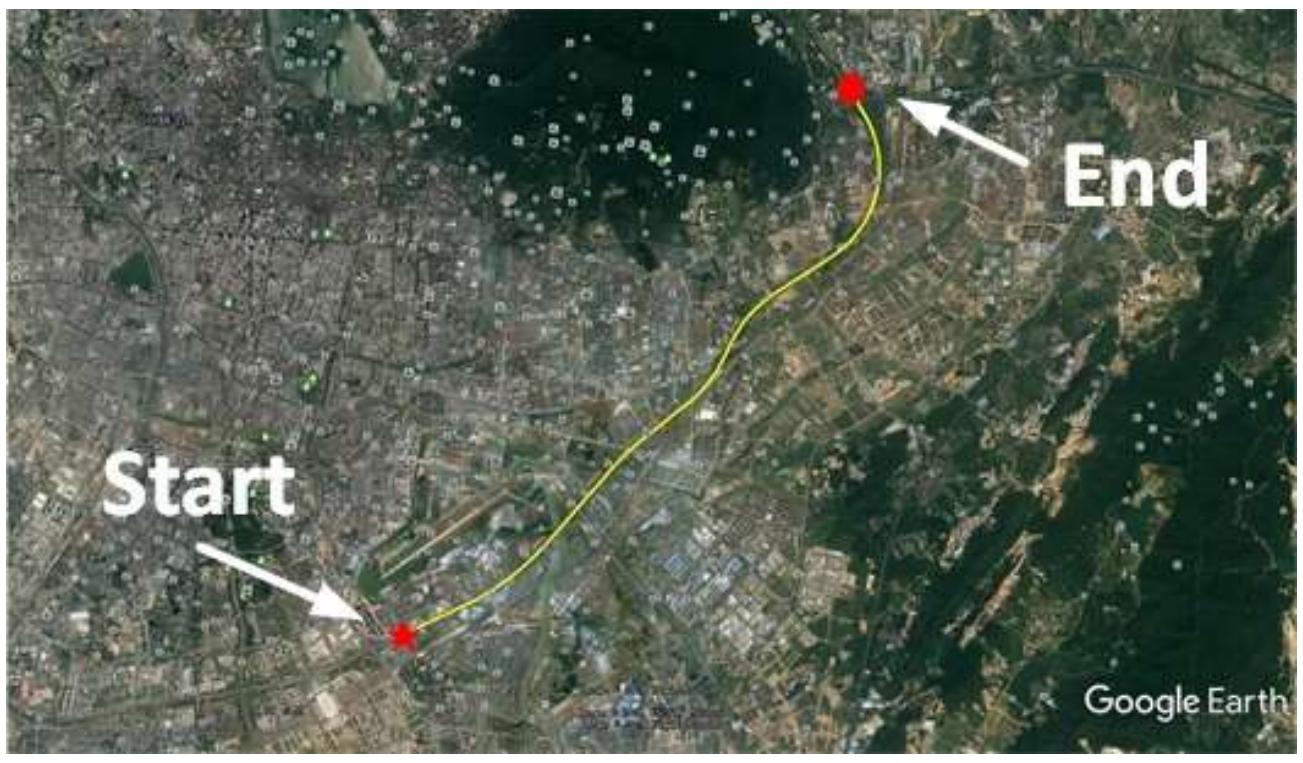

Fig. 8. Moving trajectory of the vehicle.

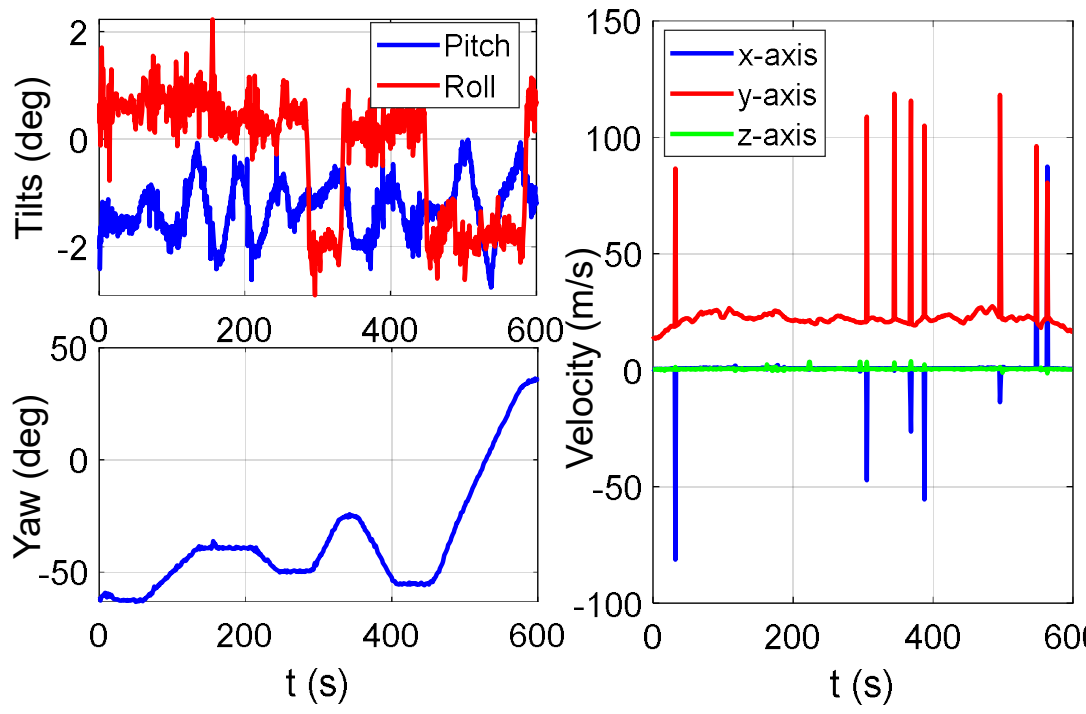

Fig. 9. Attitude and DVL outputs.

The calculated observation vectors and the reconstructed observation vectors of four schemes are depicted in the Fig. 10. It can be obviously found that the current existing work, which is scheme 1, has no ability to detect and isolate the outliers. Thus, they are contained in the calculated observation vectors, and then decrease the alignment accuracy. Compared with the scheme 1, the outliers are detected and isolated by scheme 2 effectively. The advantage of the reconstruction process for observation vectors is reflected in the alignment results, which are shown in Fig. 11 to 13.

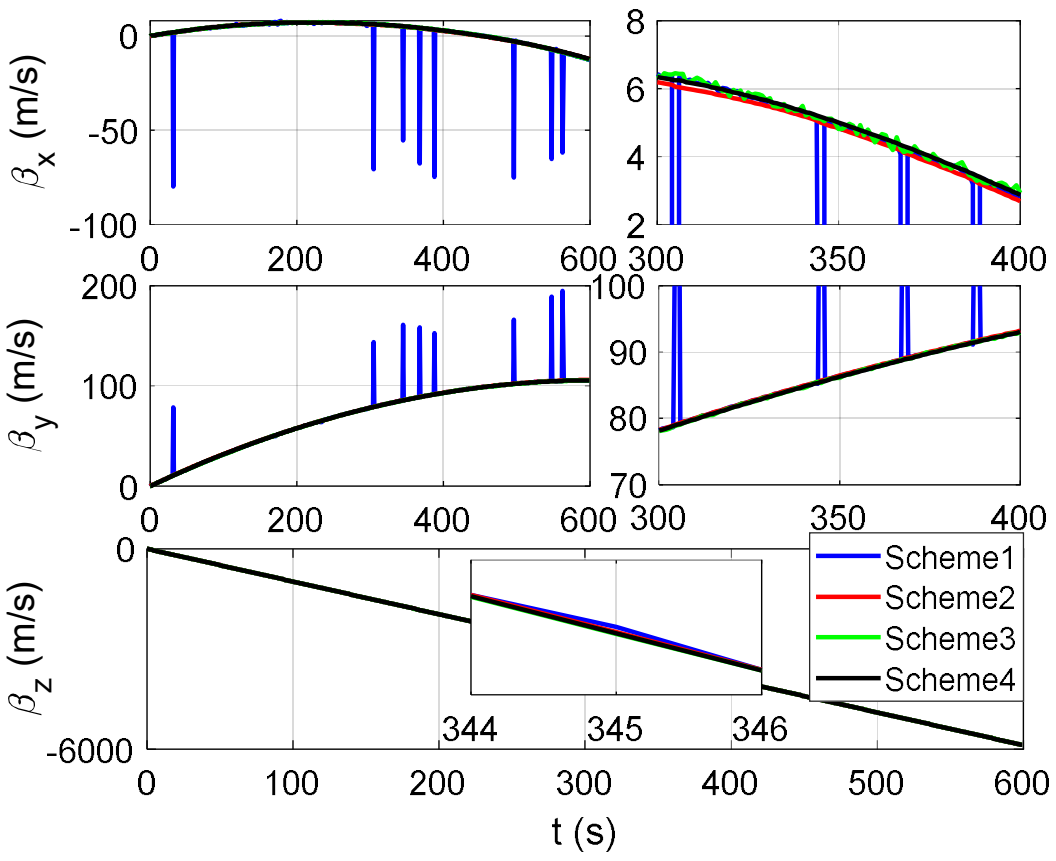

Fig. 10. Curves of the observation vectors.

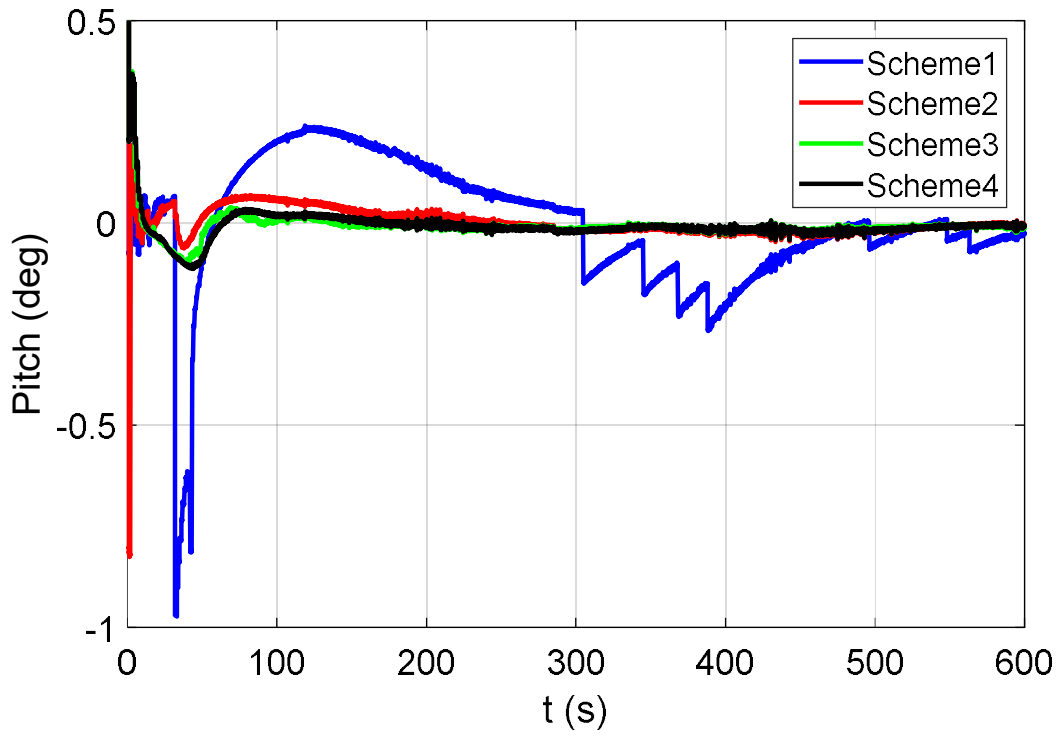

Fig. 11. Alignment errors of pitch by different schemes.

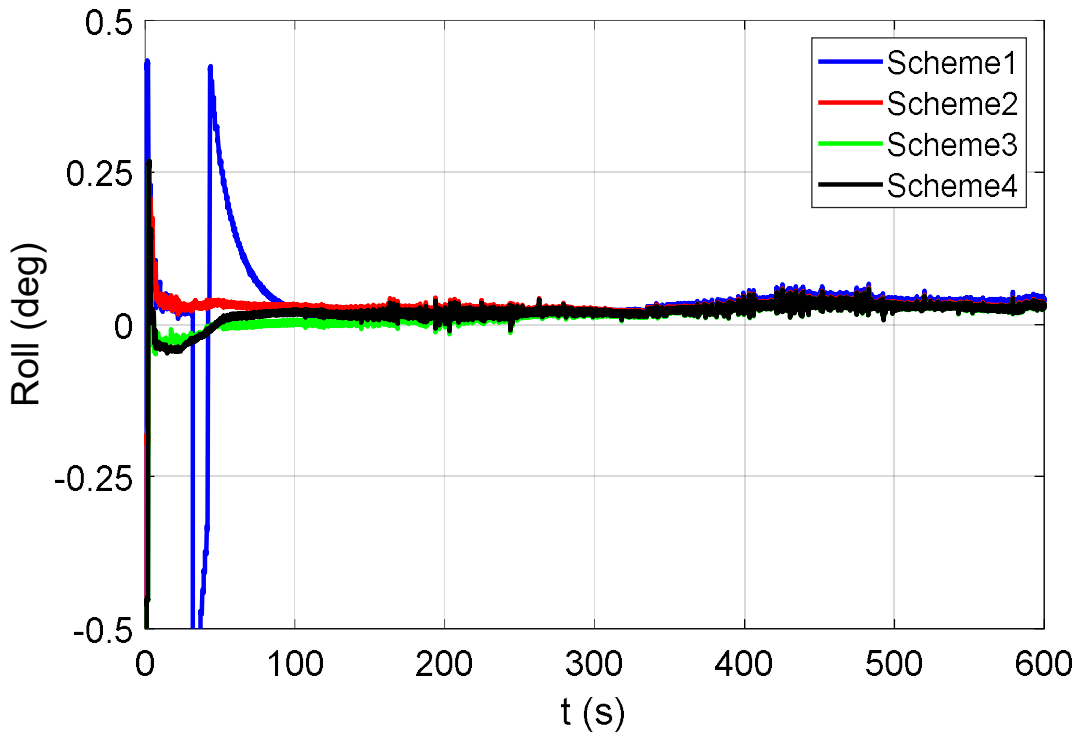

Fig. 12. Alignment errors of roll by different schemes.

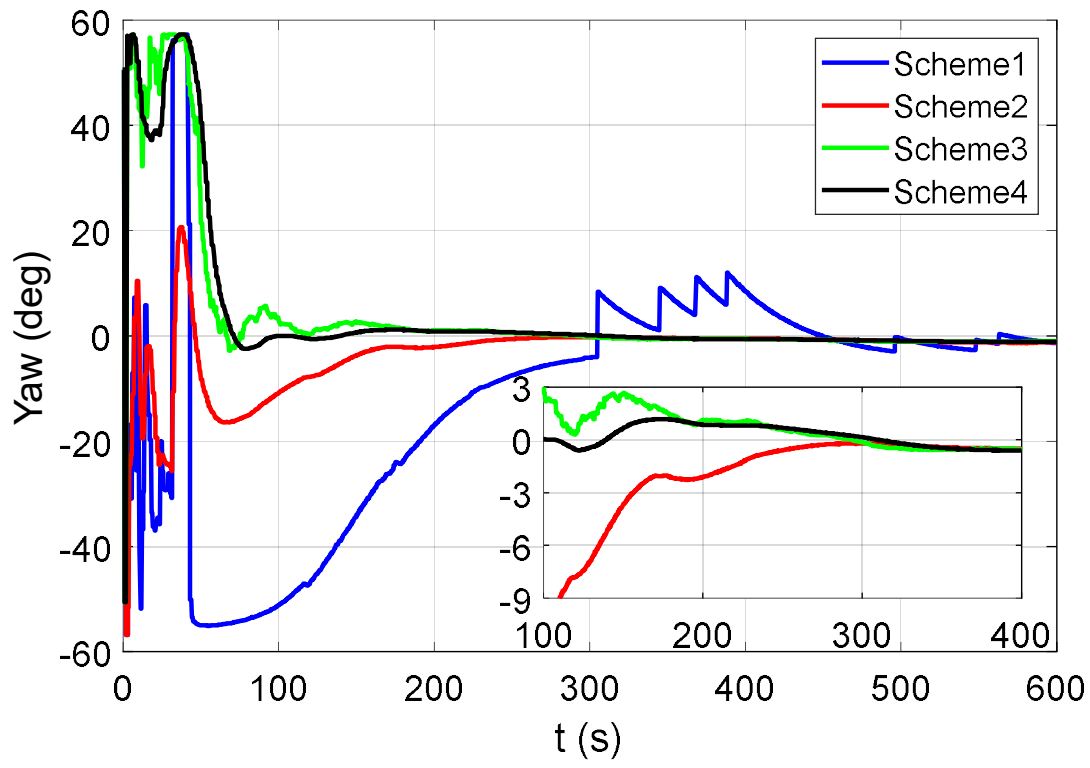

Fig. 13. Alignment errors of yaw by different schemes.

From Fig. 11 to 13, the superiority of scheme 2, which is compared with scheme 1, is obvious. The alignment results of scheme 2 show that the reconstructed method can enhance the robustness of the initial alignment. Comparing scheme 3 with 4, it shows that the reconstruction method of scheme 4, which is proposed in this paper, can smooth the alignment results. Based on this superiority, in Fig. 13, it can be found that the proposed method has faster convergence than the previous work, which is represented as scheme 3, when there are no outliers in the outputs of DVL. When alignment time lasts 300s, the errors of pitch and roll of scheme 2, 3 and 4 are less than 0.02 degree, the yaw errors of scheme 2, 3 and 4 are less than 0.5 degree. However, in Fig. 13, due to the outliers, the convergence rate of scheme 2 is slower than scheme 3 and 4, which are non-outliers contained in the outputs of DVL.

## V. Conclusion

This paper proposed a robust initial alignment method for SINS/DVL. In specific, it is shown that the reference and observation vectors are changed with the rotation of Earth. This rule reveals the essence of the apparent velocity motion. Inspired by this rule, the parameter models of the observation and reference vectors are derived. According the RKF method, the outliers contained in the calculated observation vectors are filtered out. Moreover, the more accurate observation vectors are reconstructed by the estimated parameter matrix. Through this treatment, the robust initial alignment method is developed.

Simulation and field tests studies favorably demonstrate the robustness and accuracy of the proposed approach. In specific, the proposed method can eliminate the corruption of outliers when outputs of DVL contains outliers, and smooth the alignment results compared with previous work when there is no outlier in the outputs of DVL.

## Acknowledgment

The authors thank Xiaosu Xu, Tao Zhang, Yao Li and Yiqing Yao from the School of Instrument Science and Engineering, Southeast University, China, for providing the field test.